# Entropy squeezing for a qubit with the Schrödinger cat states

Wei-Guang Wang, Xue-Qun YAN[*], and Jiu-Ming LI

Department of Physics, School of Science, Tianjin Polytechnic University, Tianjin 300387, China

State Key Laboratory of Separation, Membranes and Membrane Processes, Tianjin 300387, China

**Abstract** The manuscript investigates the entropy squeezing of a qubit (two-level atom) interacting with the cavity field in the Schrödinger cat states. The effects of the Schrödinger cat states of the field on the squeezing are examined. Our results suggest that entropy squeezing is not sensitive to the superposition features of two coherent states of the Schrödinger cat states. Furthermore, it is found that entropy squeezing occurs as the qubit initially close to its maximally coherent state, and the qubit initially in the maximally coherent state can generate optimal squeezing. This also demonstrates that the atomic coherence has a remarkable effect on the atomic squeezing. On the other hand, we also show how the entropy squeezing is related to the intensity of the initial field.

## 1  Introduction

Squeezing is a non-classical effect and also plays an important role in performing different tasks in quantum information processing such as quantum cryptography and superdense coding [1]. These quantum information tasks depend on finding the states in which squeezing can be created. In a squeezed state the quantum noise in one quadrature is below the vacuum-state level or the coherent-state level at the expense of enhanced fluctuations in the conjugate quadrature, with the product of the uncertainties in the two quadratures satisfying the uncertainty relation [2]. Specifically, an atom qubit is said to be squeezed if the quantum fluctuations of the atomic dipole are below the fundamental limit imposed by the standard uncertainty relation (SUR). The SUR is formulated in terms of the variances of the system observables. The variances containing only second-order statistical moments, however, are not an appropriate measure of the uncertainty in some cases. For example, if we use the variances as the uncertainty measure for non-Ganssian states of the radiation field, we deliberately neglect higher-order statistical moments [3]. In order to overcome the limitations of the SUR, some authors have suggested an entropic uncertainty relation [4-6]. Based on the entropic version of the uncertainty relations, a number of researchers have devoted to the atomic squeezing [7-13]. These results have shown that the entropy squeezing is more precise than the variances squeezing as a measure of the atomic squeezing. In fact, the atomic squeezing is potentially useful in various applications of quantum technology such as high-resolution spectroscopy [14], the high-precision atomic fountain clock [15], and generation of squeezed light [16].

---

[*] E-mail: yanxuequn@tjpu.edu.cn

On the other hand, in recent works it has been shown that the Schrödinger cat states（SCSs）[17], which is the superposition of two coherent states, has great potential to open up new avenues for quantum technology [18-20], including continuous-variable quantum communication [21,22], fault-tolerant quantum computation [23-26], continuous-variable teleportation [27], and quantum metrology [28-30]. Indeed, a single coherent state is a pseudo-classical state. However, a superposition of two coherent states can already display highly non-classical properties as it has been shown in some aspects [31]. There is therefore particular interest in examining the SCSs for a variety of physical properties and phenomena.

In this study, we will focus on investigating the entropy squeezing of an atom qubit when the initial state of the field is taken to be the SCSs and discuss different features of entropy squeezing in the case of the SCSs. It would also be interesting to investigate the effect of quantum coherence of the qubit on the entropy squeezing because of quantum coherence having become one of the recent research focuses in the field of both quantum mechanics and quantum technologies [32-35].

The article is organized as follows. In section 2, we give a brief review of the entropy squeezing. Section 3 is devoted to the dynamics of one photon JC model. In section 4, we investigate the properties of the atomic entropy squeezing. The paper ends with the conclusions in section 5.

## 2  Entropy Squeezing

The usual definition of squeezing based on the SUR, measures uncertainty in terms of the variances or standard deviations of the system observable. It can run into difficulties when applied to squeezing in the atom qubit. An alternative definition of squeezing is presented for this system, based on information entropy theory, which overcomes the disadvantages of the definition based on the SUR [4,5]. The basic idea of this approach is to replace the variances with the Shannon entropy as an estimator of the uncertainties associated with the measurement process. In the following we shall give a brief review about the basic concepts of the entropic uncertainty relation and entropy squeezing that will be applied later to the qubit-oscillator system.

It is well known that for a two-level system characterized by the Pauli operators $\sigma_x$, $\sigma_y$ and $\sigma_z$, with the commutation relation $\left[\sigma_x, \sigma_y\right] = 2i\sigma_z$, the SUR is given by

$$\Delta\sigma_x \Delta\sigma_y \geq \frac{1}{2}\left|\langle\sigma_z\rangle\right| \tag{1}$$

where $\Delta\sigma_\alpha = \sqrt{\langle\sigma_\alpha^2\rangle - \langle\sigma_\alpha\rangle^2}$, $(\alpha = x, y)$. The fluctuations in the components $\sigma_\alpha$ are said to be squeezed if $\sigma_\alpha$ satisfies the condition

$$V(\sigma_\alpha) = \Delta\sigma_\alpha - \sqrt{\left|\frac{\langle\sigma_z\rangle}{2}\right|} < 0. \quad \alpha = x \text{ or } y. \tag{2}$$

However, for some special atomic states, for example, the state $|\psi\rangle = \frac{1}{\sqrt{2}}(|e\rangle + |g\rangle)$ ($|e\rangle$ and $|g\rangle$ are the upper and lower eigenstates of the operator $\sigma_z$ respectively), we have $\langle\sigma_z\rangle = 0$. Hence, the uncertainty relation (1) fails to provide any useful information for atomic squeezing. Based on the entropic uncertainty relations in place of the usual SUR, the entropy squeezing, as a measure of the squeezing of the qubit, have been proposed [7].

Considering an even N-dimensional Hilbert space, an optimal uncertainty relation for sets of $N+1$ complementary observables with non-degenerate eigenvalues can be described by the inequality

$$\sum_{k=1}^{N+1} H(\sigma_k) \geq \frac{1}{2}N\ln(\frac{1}{2}N) + (1 + \frac{1}{2}N)\ln(1 + \frac{1}{2}N), \tag{3}$$

where $H(\sigma_k)$ represents the information entropy of the variable $\sigma_k$ and is defined in the following way

$$H(\sigma_k) = -\sum_{i=1}^{N} P_i(\sigma_k) \ln P_i(\sigma_k), \tag{4}$$

with

$$P_i(\sigma_k) = \langle \psi_{ki} | \rho | \psi_{ki} \rangle, \quad i = 1, 2, \ldots, N, \tag{5}$$

the probability distribution of finding the state $\rho$ in the $i$th eigenspace. $|\psi_{ki}\rangle$ is the eigenstate of the operator $\sigma_k$. The entropic uncertainty relations for a pair of observables in a finite-dimensional Hilbert space constitute the measure of uncertainty for measurement outcomes, and can be used as a general criterion for squeezing of spin system. For the two-dimensional case $N = 2$, the information entropy of the variable $\sigma_\alpha$ can be expressed as

$$H(\sigma_\alpha) = -[P_1(\sigma_\alpha) \ln P_1(\sigma_\alpha) + P_2(\sigma_\alpha) \ln P_2(\sigma_\alpha)], \quad \alpha = x, y, z. \tag{6}$$

The possibility distribution $P_i(\sigma_\alpha)$ of the operator $\sigma_\alpha$ can be obtained if the reduced density operator $\rho(t)$ of the system is known. For example, the two eigenstates of the operator $\sigma_x$ in $\sigma_z$ representation have the forms $|\psi_{x1}\rangle = (|e\rangle + |g\rangle)/\sqrt{2}$ and $|\psi_{x2}\rangle = (|e\rangle - |g\rangle)/\sqrt{2}$. From Eq. (5), we can obtain the possibility distributions of the operators $\sigma_x$ in the form

$$P_1(\sigma_x) = \langle \psi_{x1} | \rho(t) | \psi_{x1} \rangle = \frac{1}{2}\left(1 + 2\operatorname{Re}\langle e | \rho(t) | g \rangle\right) \equiv \frac{1}{2}\left(1 + 2\operatorname{Re}\rho_{eg}(t)\right) \tag{7}$$

$$P_2(\sigma_x) = \langle \psi_{x2} | \rho(t) | \psi_{x2} \rangle = \frac{1}{2}\left(1 - 2\operatorname{Re}\langle e | \rho(t) | g \rangle\right) \equiv \frac{1}{2}\left(1 - 2\operatorname{Re}\rho_{eg}(t)\right) \tag{8}$$

Similarly, we have

$$P_1(\sigma_y) = \frac{1}{2}\left(1 - 2\operatorname{Im}\langle e | \rho(t) | g \rangle\right) \equiv \frac{1}{2}\left(1 - 2\operatorname{Im}\rho_{eg}(t)\right) \tag{9}$$

$$P_2(\sigma_y) = \frac{1}{2}\left(1 + 2\operatorname{Im}\langle e | \rho(t) | g \rangle\right) \equiv \frac{1}{2}\left(1 + 2\operatorname{Im}\rho_{eg}(t)\right) \tag{10}$$

and

$$P_1(\sigma_z) = \langle g | \rho(t) | g \rangle \equiv \rho_{gg}(t), \quad P_2(\sigma_z) = \langle e | \rho(t) | e \rangle \equiv \rho_{ee}(t) \tag{11}$$

By substituting the Eqs. (7)-(11) into the formula of (6), we obtain the information entropies of the atomic operators $\sigma_x, \sigma_y$ and $\sigma_z$

$$H(\sigma_x) = -\frac{1}{2}\left(1 + 2\operatorname{Re}\rho_{eg}(t)\right) \ln\left[\frac{1}{2}\left(1 + 2\operatorname{Re}\rho_{eg}(t)\right)\right]$$

$$-\frac{1}{2}\left(1 - 2\operatorname{Re}\rho_{eg}(t)\right) \ln\left[\frac{1}{2}\left(1 - 2\operatorname{Re}\rho_{eg}(t)\right)\right] \tag{12}$$

$$H(\sigma_y) = -\frac{1}{2}\left(1 - 2\operatorname{Im}\rho_{eg}(t)\right)\ln\left[\frac{1}{2}\left(1 - 2\operatorname{Im}\rho_{eg}(t)\right)\right]$$

$$-\frac{1}{2}\left(1 + 2\operatorname{Im}\rho_{eg}(t)\right)\ln\left[\frac{1}{2}\left(1 + 2\operatorname{Im}\rho_{eg}(t)\right)\right] \tag{13}$$

$$H(\sigma_z) = -\rho_{gg}(t)\ln\left[\rho_{gg}(t)\right] - \rho_{ee}(t)\ln\left[\rho_{ee}(t)\right] \tag{14}$$

From the inequality (3), the information entropies of the operators $\sigma_x, \sigma_y$ and $\sigma_z$ satisfy the inequality

$$H(\sigma_x) + H(\sigma_y) + H(\sigma_z) \geq 2\ln 2. \tag{15}$$

If we define

$$\delta H(\sigma_\alpha) \equiv \exp[H(\sigma_\alpha)] \tag{16}$$

the inequality (15) can be expressed as

$$\delta H(\sigma_x)\delta H(\sigma_y) \geq \frac{4}{\delta H(\sigma_z)}, \tag{17}$$

which is somewhat similar to the Heisenberg uncertainty production of a spin system. It shows that there is impossibility of simultaneously having complete information about both observable $\sigma_x$ and $\sigma_y$. The component $\sigma_\alpha$ are said to be squeezed in entropy if the Shannon information entropy satisfies

$$E(\sigma_\alpha) = \delta H(\sigma_\alpha) - \frac{2}{\sqrt{\delta H(\sigma_z)}} < 0, \quad \alpha = x, y. \tag{18}$$

where $E(\sigma_\alpha)$ are the entropy squeezing factors. In the following analysis, we shall use the equation (18) to examine the time evolution of the entropy squeezing factors and the atomic squeezing of the system under consideration.

## 3 Model and solution

By considering the resonance condition, in the interaction picture, the interaction of a single-model quantized field with a single atom qubit can be described by the Hamiltonian ($\hbar = 1$)

$$H = \lambda(\sigma_+ a + \sigma_- a^+) \tag{19}$$

where the operators $\sigma_\pm$ are the qubit ladder operators $\sigma_+ = |e\rangle\langle g|$ and $\sigma_- = |g\rangle\langle e|$. The two

operators $\sigma_+ a$ and $\sigma_- a^+$ correspond respectively to a transition from the lower level $|g\rangle$ to the upper level $|e\rangle$ together with the annihilation of a photon, and the transition from $|e\rangle$ to $|g\rangle$ together with the emission of a photon. The corresponding evolution unitary operator for an interaction time $t$ reads

$$U(t) = \exp(-iHt) = \exp(-iG\tau) \qquad (20)$$

where we defined $G = \sigma_+ a + \sigma_- a^+$ and $\tau = \lambda t$.

In our treatment we assume that the initial state of the qubit and cavity to be a pure product state, thus the density operator for the system at initial moment is given by

$$\rho(0) = |\Psi_0\rangle\langle\Psi_0| \qquad (21)$$

with $|\Psi_0\rangle = |\psi_a\rangle \otimes |\psi_f\rangle$. In particular, the qubit at time $t=0$ is prepared in a pure superposition of the upper and lower states of the system,

$$|\psi_a\rangle = \cos\frac{\theta}{2}|e\rangle + e^{-i\phi}\sin\frac{\theta}{2}|g\rangle \qquad (22)$$

while the bosonic fields are prepared in the SCSs in a single mode,

$$|\psi_f\rangle = N(|\alpha\rangle + e^{i\rho_c}|-\alpha\rangle) \qquad (23)$$

where $N = \left(2 + 2e^{-2|\alpha|^2}\cos\rho_c\right)^{-\frac{1}{2}}$ and the coherent state of the field is

$$|\alpha\rangle = e^{-\frac{1}{2}|\alpha|^2}\sum_{n=0}^{\infty}\frac{\alpha^n}{\sqrt{n!}}|n\rangle = \sum_n P_n|n\rangle \qquad (24)$$

Here $\alpha = \sqrt{\bar{n}}e^{i\beta}$, where the mean photon number $\bar{n}$ shows the intensity of the field and $\beta$ is its phase; and $P_n = e^{-\frac{1}{2}|\alpha|^2}\frac{\alpha^n}{\sqrt{n!}}$. The evolution of the total system reads

$$\rho(\tau) = U(\tau)\rho(0)U^+(\tau) \qquad (25)$$

and upon tracing the evolved state over the bosonic field, we obtain the states

$$\rho_a(\tau) = Tr_f\left[U(\tau)\rho(0)U^+(\tau)\right] \qquad (26)$$

which describe the qubit state at time $t$.

In details, the reduced qubit density operator reads in the basis $\{|e\rangle, |g\rangle\}$ as

$$\rho_a(\tau) = \begin{pmatrix} \rho_{ee} & \rho_{eg} \\ \rho_{ge} & \rho_{gg} \end{pmatrix} \tag{27}$$

where the matrix elements are

$$\rho_{ee} = N^2 \sum_n |P_n|^2 \left[ 2(1+\cos\rho_c \cdot (-1)^n) \cdot \cos^2\frac{\theta}{2} \cdot \cos^2(\tau\sqrt{n+1}) \right.$$
$$+ 2\left(1+\cos\rho_c \cdot (-1)^n\right) \cdot \sin^2\frac{\theta}{2} \cdot \sin^2(\tau\sqrt{n})$$
$$\left. -(-1)^n \cdot \sqrt{\frac{\bar{n}}{n+1}} \cdot \sin\rho_c \cdot \sin\theta \cdot \cos(\phi-\beta) \cdot \sin(2\tau\sqrt{n+1}) \right] \tag{28}$$

$$\rho_{gg} = 1 - \rho_{ee} \tag{29}$$

$$\rho_{eg} = N^2 \cdot \sin\theta \sum_{n=0}^{\infty} \left(1+(-1)^n \cdot \cos\rho_c\right)\left[ e^{i\phi} \cdot |P_n|^2 \cdot \cos(\tau\sqrt{n+1}) \cdot \cos(\tau\sqrt{n}) \right.$$
$$\left. + e^{-i\phi} \cdot P_{n+1} \cdot P_{n-1}^* \sin(\tau\sqrt{n+1}) \cdot \sin(\tau\sqrt{n}) \right]$$
$$-2N^2 \cdot \sin\rho_c \sum_{n=0}^{\infty} (-1)^n \left[ \cos^2\frac{\theta}{2} \cdot P_n \cdot P_{n-1}^* \cos(\tau\sqrt{n+1}) \cdot \sin(\tau\sqrt{n}) \right.$$
$$\left. + \sin^2\frac{\theta}{2} \cdot P_{n+1} \cdot P_n^* \cdot \sin(\tau\sqrt{n+1}) \cdot \cos(\tau\sqrt{n}) \right] \tag{30}$$

$$\rho_{ge} = \rho_{eg}^* \tag{31}$$

Using the above equations, we can study the time evolution of the entropy squeezing factors of the observables $\sigma_x$ and $\sigma_y$. It will be done in the next section.

## 4 Numerical results and discussion

We now investigate numerically the dynamical properties of entropy squeezing factors and the entropy squeezing by considering a atom qubit with the SCSs for various parameter regimes. The numerical results of evolution of the entropy squeezing factors $E(\sigma_x)$, $E(\sigma_y)$ are shown in Figs. 1 in the cases where the qubit is initially in the upper state ($\theta = 0$) and the field is in the SCSs with mean photon number $\bar{n} = 25$, $\rho_c = \pi/6$ and phase $\beta = 0$. In Figs. 1(a) and (b), we depict long-time behavior of the entropy squeezing factors $E(\sigma_x)$ and $E(\sigma_y)$ respectively. It is seen that there is the entropy squeezing in both $E(\sigma_x)$ and $E(\sigma_y)$ for very early times. In order to obtain more precise information about the squeezing, we plot in Fig. 1(c) a short-time behavior of the entropy squeezing factors. It is obvious from this figure that there exist a sequence of collapses and revivals and the amplitudes of the revivals decrease as time goes by, and $E(\sigma_x)$ and $E(\sigma_y)$ fluctuate oppositely. Also, one can find entropy squeezing in both variables $\sigma_x$ and $\sigma_y$ at some intervals of time,

To examine the effect of the SCSs of the field on the entropy squeezing, Fig. 2 is plotted for parameter $\rho_c = 0$. Comparing Fig. 1(c) and Fig. 2 one can see that both figures almost coincide with each other. It leads us to conclude that entropy squeezing factors $E(\sigma_x)$, $E(\sigma_y)$ are independent of the parameter $\rho_c$. And we conjecture that the entropy squeezing is not sensitive to the superposition features of the two coherent states of the SCSs.

More recently based on the perspective of resource theory, quantum coherence, which arises from the superposition principle of quantum states, have become one of the research focuses in the field of both quantum mechanics and quantum information science. In order to investigate the effect of quantum coherence of the qubit on the entropy squeezing, it is instructive to examine the time evolution of the entropy squeezing factors for different atomic states (22) of different parameters $\theta$. In Figs. 3, we plot the entropy squeezing factors $E(\sigma_x)$, $E(\sigma_y)$ against the scaled time $\tau$, for parameters $\bar{n} = 25$, $\beta = \pi/4$, $\rho_c = \pi/6$, $\phi = \pi/2$ and with different values of the parameters $\theta$. It is observed from these figures that the time evolution of the entropy squeezing factors can be drastically different when the parameter $\theta$ changes. We notice that the entropy squeezing occurs when the qubit is initially close to the maximally coherent state (see Fig. 3(c), for instance). In Fig. 1(c), for the maximally coherent state of qubit, $|\psi_a\rangle = \frac{1}{\sqrt{2}}(|e\rangle + i|g\rangle)$, it is easy to see that the $E(\sigma_x)$ and $E(\sigma_y)$ exhibit optimal entropy squeezing. In addition, we also note in Fig. 3(a) and (e) that no entropy squeezing occurs when the qubit is initially in an incoherent state (the upper state $|e\rangle$ or the lower state $|g\rangle$), and the time evolutions of the entropy squeezing factors both $E(\sigma_x)$ and $E(\sigma_y)$ become agreement with each other. This emphasizes the fact that the atomic coherence has a remarkable effect on the entropy squeezing.

In order to have a clearer understanding of the effects of the intensity of the field on the entropy squeezing, the entropy squeezing factors both $E(\sigma_x)$ and $E(\sigma_y)$ are plotted as a function of the scaled time $\tau$ in Figs. 4. We can see from Figs. 4(a) and (b) that there is squeezing in both $E(\sigma_x)$ and $E(\sigma_y)$ at different intervals of time for early times when the initial field is stronger (e.g., $\bar{n}=15$ and 5). Also one can see that the number of squeezing decrease, but the time interval of the squeezing increase, as $\bar{n}$ decreases. In Figs. 4(c) and (d) we see, the situation being rather different, that for weak initial fields ($\bar{n}<1$), the $E(\sigma_x)$ and $E(\sigma_y)$ predict no entropy squeezing on the variable $\sigma_x$, but entropy squeezing occurs only for the variable $\sigma_y$. In Fig. 4(d), for very weak fields one can see a periodical oscillation of the entropy squeezing factors, and there is a optimal entropy squeezing on the variable $\sigma_y$. It is readily found that the optimal entropy squeezing is attained in the eigenstate of the atomic operator $\sigma_y$.

## 5   Conclusions

In the present work, we have investigated the features on entropy squeezing for a qubit with the SCSs. Our results have shown that for the qubit there exists a rich feature of entropy squeezing in the case of the SCSs.

In order to elucidate the role of SCSs, we have examined the influence of different superposition of two coherent states of the cat state on the entropy squeezing. We have also shown that the entropy squeezing is not sensitive to the superposition features of two coherent states of the SCSs. Moreover, we have found that the atomic coherence has a remarkable effect on the entropy squeezing, and the atom qubit initially in the maximally coherent state can generate optimal entropy squeezing.

To study the effects of the intensity of the initial field on the entropy squeezing, we have examined the entropy squeezing factors as a function of time for different values of the mean photon number $\bar{n}$. We have found that there is squeezing on both variables $\sigma_x$ and $\sigma_y$ at different intervals of time when the initial fields are strong, while for weak initial fields ($\bar{n} < 1$) entropy squeezing occurs only for the variable $\sigma_y$ and for very weak fields (e.g., $\bar{n} = 0.05$) there is a periodic oscillation of the entropy squeezing factors.

As a future work, we intend to further establish the intrinsic connection between quantum coherence of atom qubit and entropy squeezing from the perspective of resource theory. We believe that these studies will also prove useful in the continued efforts to understand the complex physics in the quantum uncertainty relation.

## Acknowledgment

This work is supported by the Science and Technology Plans of Tianjin (No. Z2-201554)

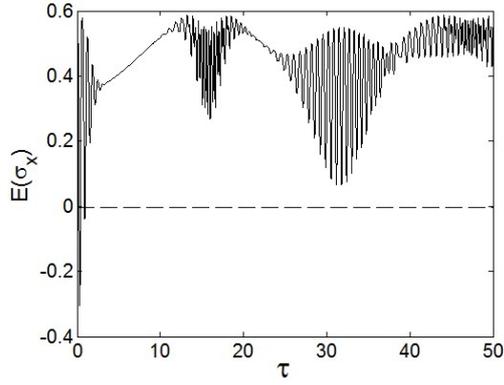

**(a)**

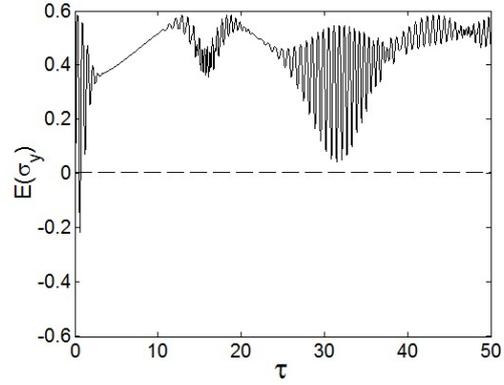

**(b)**

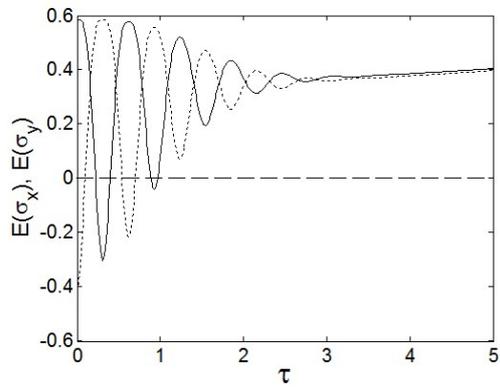

**(c)**

Fig. 1. The time evolution of the entropy squeezing factors $E(\sigma_x)$, $E(\sigma_y)$ of a qubit in the JC model, for parameters $\beta = \pi/4$, $\theta = \pi/2$, $\rho_c = \pi/6$, $\phi = \pi/2$ and $\bar{n} = 25$. (a) The entropy squeezing factor $E(\sigma_x)$ against scaled time $\tau$; (b) the entropy squeezing factor $E(\sigma_y)$ against scaled time $\tau$; (c) the entropy squeezing factors $E(\sigma_x)$ (solid line) and $E(\sigma_y)$ (dotted line) against scaled time $\tau$.

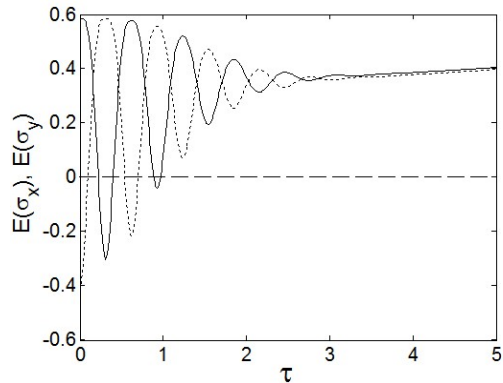

Fig. 2. Same as figure 1(c), but with $\rho_c = 0$ instead.

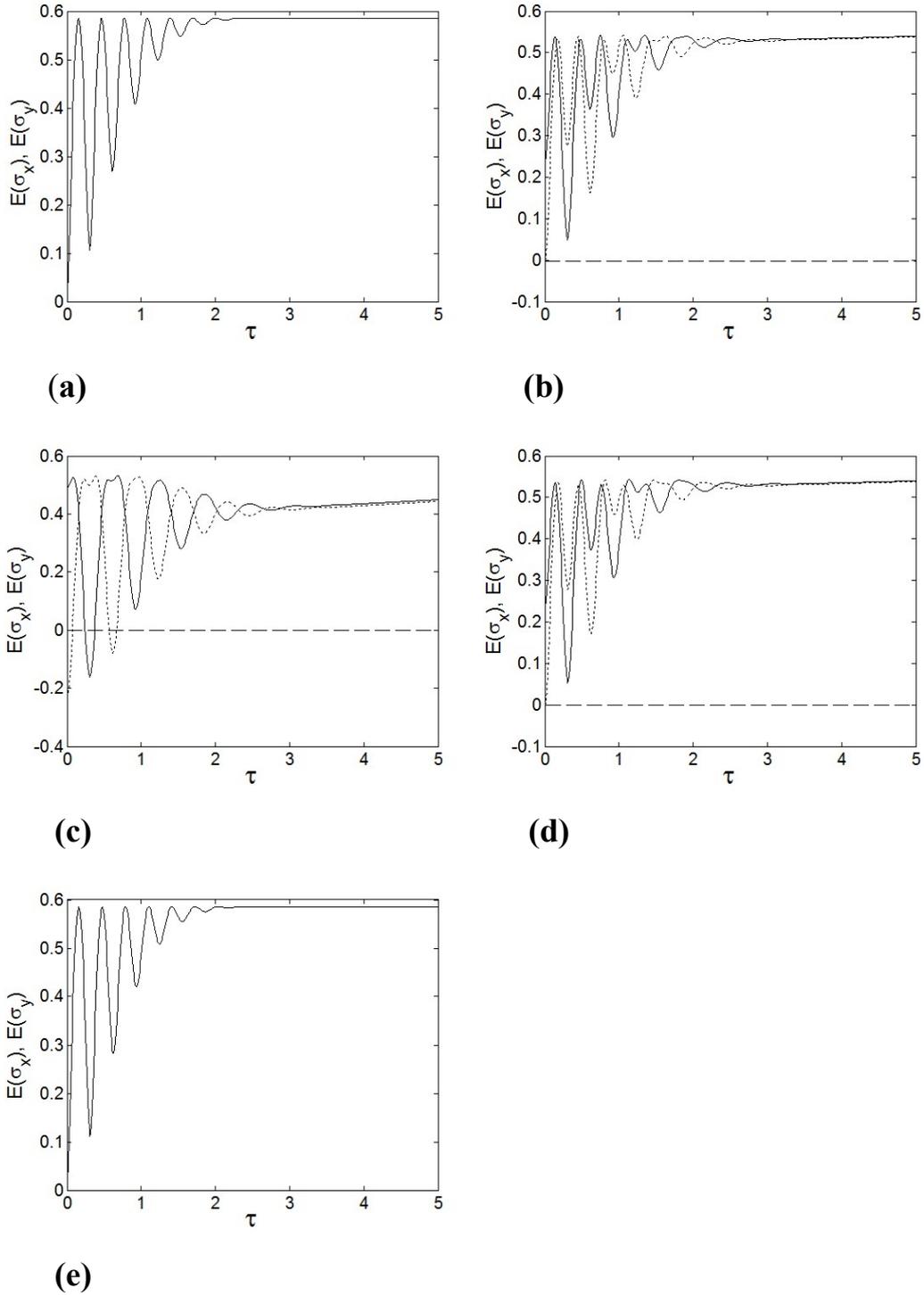

(a)

(b)

(c)

(d)

(e)

Fig. 3. The time evolution of the entropy squeezing factors $E(\sigma_x)$ (solid line), $E(\sigma_y)$ (dotted line) of a qubit in the JC model for parameters $\beta = \pi/4$, $\rho_c = \pi/6$, $\phi = \pi/2$, $\bar{n} = 25$ and with different values of the parameter $\theta$ where $\theta = 0$ in figure 3(a), $\theta = \pi/6$ in figure 3(b), $\theta = 2\pi/6$ in figure 3(c), $\theta = 5\pi/6$ in figure 3(d), $\theta = \pi$ in figure 3(e).

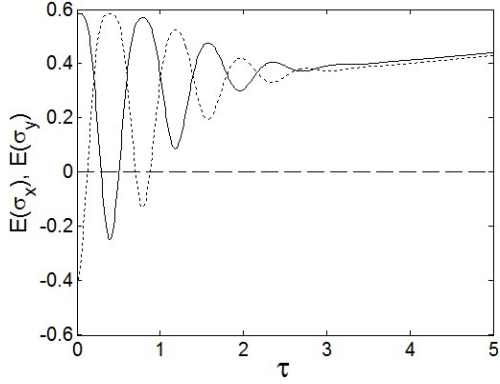
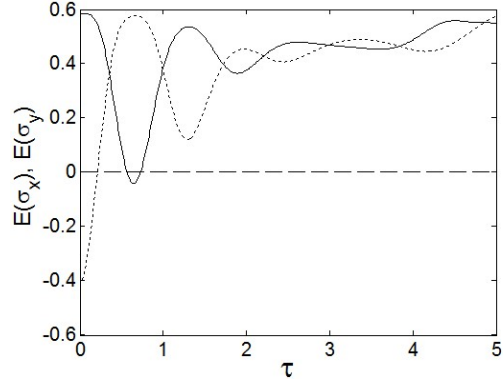

(a)    (b)

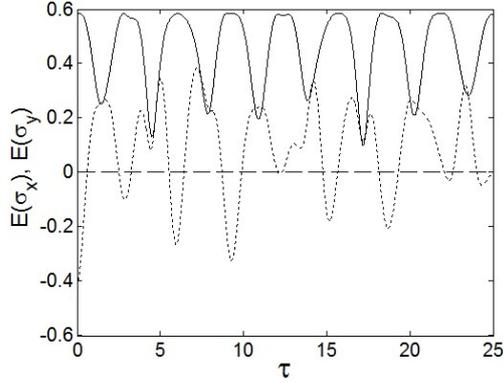
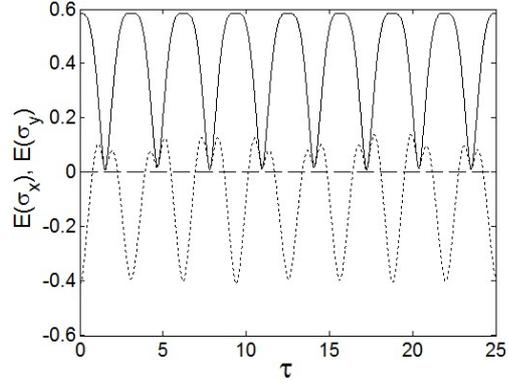

(c)    (d)

Fig. 4. The time evolution of the entropy squeezing factors $E(\sigma_x)$ (solid line), $E(\sigma_y)$ (dotted line) of a qubit in the JC model for parameters $\beta = \pi/4$, $\theta = \pi/2$, $\rho_c = \pi/6$, $\phi = \pi/2$, and with different values of the mean photon number $\bar{n}$ where $\bar{n} = 15$ in figure 4(a), $\bar{n} = 5$ in figure 4(b), $\bar{n} = 0.5$ in figure 4(c), $\bar{n} = 0.05$ in figure 4(d).